\documentstyle{cupconf}

\ifoldfss
\else
  \ifnfssone
    \newmathalphabet{\mathit}
      \addtoversion{normal}{\mathit}{cmr}{m}{it}
      \addtoversion{bold}{\mathit}{cmr}{bx}{it}
    \newmathalphabet{\mathcal}
      \addtoversion{normal}{\mathcal}{cmsy}{m}{n}
    \else
    \ifnfsstwo
    \fi
  \fi
\fi

\def\hexnumber#1{\ifcase#1 0\or1\or2\or3\or4\or5\or6\or7\or8\or9\or
 A\or B\or C\or D\or E\or F\fi }

\makeatletter
\ifx\CUP@mtlplain@loaded\undefined
\else
\fi
\makeatother
 \makeatletter
 \ifx\CUP@mtlplain@loaded\undefined
   \font\tenbmi=cmmib10 at 10pt
   \font\sevenbmi=cmmib10 at 7pt
   \font\fivebmi=cmmib10 at 5pt

   \newfam\bmifam
   \textfont\bmifam=\tenbmi
   \scriptfont\bmifam=\sevenbmi
   \scriptscriptfont\bmifam=\fivebmi
   
 \fi
 \makeatother
\ifnfsstwo

\fi
\ifnfssone

\fi
\ifoldfss

\fi

\mathchardef\varLambda="0103

\makeatletter
\ifx\CUP@mtlplain@loaded\undefined
\else
\fi
\makeatother
\makeatletter
\ifx\CUP@mtlplain@loaded\undefined
  \font\tenbms=cmbsy10
  \font\sevenbms=cmbsy10 at 7pt
  \font\fivebms=cmbsy10 at 5pt
  \newfam\bmsfam
  \textfont\bmsfam=\tenbms
  \scriptfont\bmsfam=\sevenbms
  \scriptscriptfont\bmsfam=\fivebms

  \edef\bsy@{\hexnumber\bmsfam}
  \mathchardef\bnabla="0\bsy@72
\fi
\makeatother
\title[Galactic Black Hole X-ray Binaries]{Black Holes in our Galaxy:
Observations}

\author[P. A. Charles]%
{P\ls H\ls I\ls L\ns C\ls H\ls A\ls R\ls L\ls E\ls S\ls}

\affiliation{Department of Astrophysics, Oxford University,
Keble Road, Oxford OX1 3RH, UK\\}

\begin{document}
\ifnfssone
\else
  \ifnfsstwo
  \else
    \ifoldfss
      \let\mathcal\cal
      \let\mathrm\rm
      \let\mathsf\sf
    \fi
  \fi
\fi

\maketitle

\begin{abstract}

This paper reviews the X-ray, optical, radio and IR observations of galactic X-ray
binaries suspected to contain black hole compact objects, with particular
emphasis on the supporting dynamical evidence.

\end{abstract}

\firstsection 

\section{Introduction}

From just one object (Cyg X-1) 25 years ago, the number of strong 
black-hole candidates in the Galaxy has increased dramatically in
the last decade.  This is due to two main factors: (i) the provision
of all-sky monitoring capabilities on the current generation of X-ray
observatories, and (ii) the discovery of the class of low-mass X-ray
binary (LMXB) known as {\it soft X-ray transients} (SXT).  These
are short-period
(5hrs -- 6days) binaries with mostly K--M type secondaries, and many
similarities with dwarf novae.  Their
rare, dramatic X-ray outbursts (typically separated by decades) can reach
very high luminosities ($>10^{38}$erg s$^{-1}$) and are considered to
be accretion disc instability events.  Remarkably, a very high fraction
of SXTs are black-hole candidates.  This is largely because they are the
only LMXBs for which accurate mass determinations are possible, through
the ability to undertake dynamical studies of the secondary star during
the long periods of quiescence.  While Cyg X-1 has been joined by other
high-mass X-ray binaries (HMXB) also believed to contain black-holes,
they all suffer from the serious limitation that an accurate mass is
required for the early-type mass donor before it is possible to precisely
determine the compact object mass.  Given their complex binary 
evolutionary path this is difficult to obtain.

It has been more than 30 years since the first X-ray binary was optically 
identified (Sco X-1), but a detailed knowledge of their binary parameters 
only started to come in the 1970s with the identification and study of 
Cyg X-1, Her X-1 and Cen X-3.  However, these all have early-type companions, 
observable in spite of the presence in the binary of luminous X-ray 
emission (for a review see van Paradijs \& McClintock 1995).  
Apart from Cyg X-1, most of these high-mass X-ray binaries
(HMXRBs) had 
pulsating neutron star compact objects, thereby providing the potential 
for a full solution of the binary parameters since they were essentially 
double-lined spectroscopic binaries.  From this has come the detailed 
dynamical mass measurements of neutron star systems which have recently 
been collated by Thorsett \& Chakrabarty (1998), showing that they are 
all consistent with a mass of 1.35$\pm$0.04M$_\odot$.

However, when HMXRBs are suspected of harbouring much more massive 
compact objects (as is the case for Cyg X-1), the mass measurement 
process runs into difficulties.  By definition, there will be no dynamic 
features (such as pulsations) associated with the compact object that can 
be observed.  Hence all mass information must come from the mass-losing 
companion, and the mass of the compact object cannot be determined unless 
the companion's mass is accurately known.

The situation for low-mass X-ray binaries (LMXBs), such as Sco X-1, is
completely different, in that 
their short orbital periods require their companion stars to be of low 
mass.   This can be demonstrated quite simply as follows (see King 1988).  
Since these are interacting binaries in which the companion fills 
its Roche lobe, then we may employ the useful Paczynski (1971) relation 

\begin{equation}
R_2/a = 0.46(1+q)^{-1/3} 
\end{equation}

where the mass ratio $q = M_X/M_2$.  Combining
this with Kepler's 3rd Law yields the well-known result that the mean
density, $\rho$ (in g~cm$^{-3}$) of the secondary, 

\begin{equation}
{\rho} = 110/P^2_{hr}   
\end{equation}

And if 
these stars are on or close to the lower main sequence, then $M_2=R_2$ 
and hence $M_2=0.11P_{hr}$.  Therefore short period X-ray binaries must 
be LMXBs and so the companion star will be faint.  The major observational
problem with this is that the optical light 
will then be dominated by reprocessed X-radiation from the disc (or heated 
face of the companion star; see van Paradijs \& McClintock 1994).  This 
is why the optical spectra of LMXBs are hot, blue continua (U--B 
typically -1) with superposed broad hydrogen and helium emission lines, 
the velocities of which indicate that they largely arise in the inner 
disc region, thereby denying us access to the dynamical information that 
is essential if accurate masses are to be determined.  
Hence, the evidence for the nature of the compact object in most bright 
LMXBs has come from indirect means, usually X-ray bursting behaviour 
(as few are X-ray pulsars) or the fast flickering first seen in Cyg X-1 
(and hence used as a suggestion of the presence of a black hole).
[Note, however, that while it is useful to employ the Paczynski relation 
in this way, it is only valid for $q>1$, and there is a more accurate 
algorithm due to Eggleton (1983) which is valid for all $q$.]

To make real progress in determining the nature of compact objects in our 
Galaxy requires dynamical mass measurements of the type hitherto employed 
on neutron stars in HMXRBs.  But without velocity information associated 
with the compact object, all that can be measured (in the case of 
Cyg X-1, and the other two HMXRBs suspected of harbouring black holes, 
LMC X-1 and LMC X-3) 
is the mass function 

\begin{equation}
f(M) = {{PK^3}\over {2{\pi}G}} = {{M^3_X{\sin}^3i}\over{(M_X+M_2)^2}}
\end{equation}

where $P$ is the orbital period and $K$ is the radial velocity amplitude.
And 
since $M_2{\geq}M_X$, then $M_X$ is not accurately known because $M_2$ has 
a wide range of uncertainty ($\sim$12--20M$_\odot$) given the unusual 
evolutionary history of the binary.  The compact object in Cyg X-1 almost 
certainly is a black hole, but an accurate mass determination is not 
possible from the available data which simply constrain it to be 
$>$3.8M$_\odot$ (Herrero et al 1995).  
This is close to the canonical maximum mass of a 
neutron star, based on the oft-quoted Rhoades-Ruffini Theorem (1974).  
However, there are a number of assumptions built into this which need 
careful examination in the light of the masses of the compact objects 
reviewed here (see e.g. Miller 1998 and Miller et al 1998).

For the LMXBs we clearly need to find systems in 
which the companion star {\it is} visible, which requires sources where 
the X-ray emission switches off for some reason.  This is the basis of 
the new field of study of the {\it soft X-ray transients}, hereafter 
SXTs (and 
sometimes referred to as {\it X-ray novae}).  Remarkably, of the $\sim$23 
currently known, only 6 (i.e. $\sim$25\%) are confirmed neutron star 
systems (they display type I X-ray bursts), the remainder are all 
black-hole candidates, the highest fraction of any class of X-ray source.

\section{X-ray/Optical Behaviour}

\subsection{Outburst}

The SXTs typically outburst every $\sim$10--20 years.  The first one (Cen
X-2) was found by early rocket flights (Harries et al 1967), but the
prototype of the class (due to its proximity and detailed multi-wavelength
study) is widely considered to be A0620-00 (Nova Mon 1975), for several
months the brightest X-ray source in the sky and peaking at 11th mag in the
optical (Elvis et al 1975; see figure 1, taken from Kuulkers 1998). Their
light curves tend to show a fast rise followed by an exponential decay (see
Chen et al 1997 for a compendium of all SXT light curves), the optical
amplitude of which has been shown by Shahbaz \& Kuulkers (1998) to be
related to the orbital period, and the precise form of the decay is related
to the peak X-ray luminosity at outburst (King \& Ritter 1997; Shahbaz et al
1998a).

\begin{figure}[t]
\begin{center}
\begin{picture}(100,250)(50,30)
\put(0,0){\includegraphics{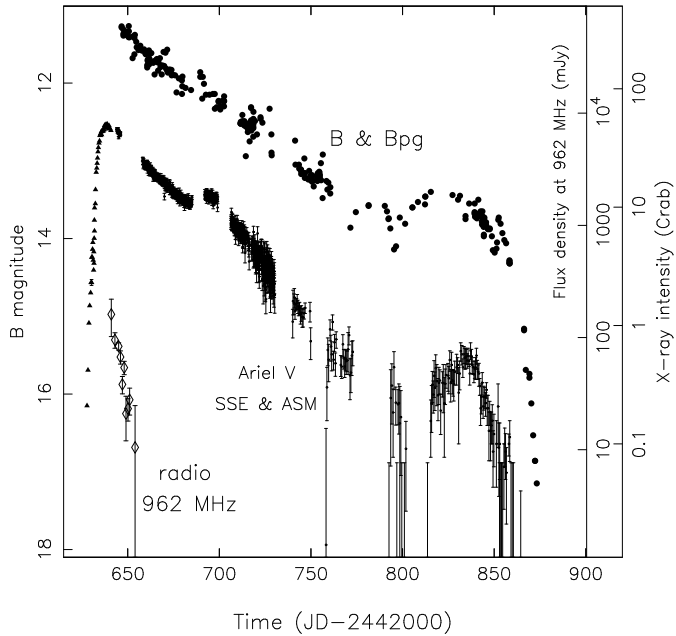}}
\noindent
\end{picture}
\caption[{\bf Figure 1}] {Optical, X-ray and radio outburst light curves 
of the prototype SXT A0620-00 (=Nova Mon 1975), adapted from Kuulkers (1998).}

\label{fig1}
\end{center}
\end{figure}

It takes $\sim$1 year for SXTs to reach optical/X-ray quiescence after an
outburst, but note that there have been subsequent {\it mini}-outbursts in
some systems (e.g. GRO~J0422+32, Kuulkers 1998 and references therein) and
erratic re-brightenings in others (e.g. GRO~J1655-40).  The observed
properties of the 9 SXTs for which full dynamical analyses have been
performed are summarised in table 1 and listed in order of orbital period,
apart from separating the two SXTs which display a much earlier spectral
type companion star.

At the time of outburst most (but not all) exhibit {\it ultra-soft} X-ray 
spectra with black-body colour temperatures of $kT{\sim}0.5-1$keV 
superposed on a hard power-law extending to much higher energies (see 
Tanaka \& Lewin 1995).  It is this characteristic that gives the SXTs 
their name, and effectively distinguishes them from the much harder Be 
X-ray transients that appear to be almost exclusively long-period neutron 
star systems.  [Note also that the ultra-soft X-ray spectrum of SXTs is very 
different from the {\it super-soft} designation applied (mostly) to the 
(very) much cooler accreting white dwarf systems in the LMC and SMC (see
Kahabka \& van den Heuvel 1997).]

Additionally the SXTs (e.g. GS2023+338) can show extremely erratic
variability which is very similar to that displayed by Cyg X-1.  Hence the
X-ray spectrum and variability are used as key discriminators to hunt for
black holes.  However, it must be noted that, in certain circumstances,
neutron star systems can mimic these properties (e.g. Cir X-1 and X0331+53),
and so we must use only dynamical evidence in the final analysis as to the
nature of the compact objects (McClintock 1991).

\begin{table}
\begin{center}
\caption{Optical/IR Properties of Soft X-ray Transients}
\begin{tabular}{lcccccccc} 
\hline
{\em Source} & {\em Outbursts} & {\em P} & {\em Sp.} &  
{\em E$_{B-V}$} & {\em V} & {\em K} & 
{\em v$\sin$i} & {\em K$_2$}\\  
 &  & {\em (hrs)} & {\em Type} &  &{\em (quiesc)} & & 
{\em (km/s)} & {\em (km/s)} \\ \hline

J0422+32 & 1992 & 5.1 & M2V & 0.3   & 22 & 16.2 & $\leq$80 &  381\\
A0620-00 & 1917,75 & 7.8 & K5V &   0.35 & 18.3 &  6 & 83 & 433\\
GS2000+25 & 1988 & 8.3 & K5V & 1.5  & 21.5 & 17  & 86 & 518\\
GRS1124-68 & 1991 & 10.4 & K0-4V &  0.29 & 20.5 & 16.9 & 106 & 399\\
H1705-25 & 1977 & 12.5 & K & 0.5 & 21.5 & - & $\leq$79  & 448\\
Cen X-4 & 1969,79 & 15.1 & K7IV &   0.1  & 18.4 & 15.0 & 45 & 146\\
V404 Cyg & 1938,56,89 & 155.3 & K0IV &  1 & 18.4 & 12.5 & 39 & 208.5\\
 & & & & & & & & \\
4U1543-47 & 1971,83,92 & 27.0 & A2V & 0.5 & 16.6 & - & - & 124\\
J1655-40 & 1994+ & 62.9 & F3-6IV & 1.3 & 17.2 & - & - & 228\\
\end{tabular}
\end{center}
\end{table}

\subsection{Quiescence}

Even in quiescence, optical studies (see section 3) show that mass 
transfer continues in the SXTs, and indeed many have been detected by 
X-ray observatories (Einstein, EXOSAT, ROSAT) as very weak sources (e.g.
Verbunt et al 1994).  
However, the observed luminosities are substantially lower than expected 
for the continuing accretion rate, and this has led various groups
(see e.g. Abramowicz et al 1995; 
Narayan et al 1997a) to propose that {\it advective accretion} is taking place.  
The inner disc at low accretion rates evaporates due to 
the X-radiation into a very hot low density corona.  Such hot gas cannot 
radiate efficiently and transports most of its thermal energy onto the 
compact object (the advection process).  (Such models can also account 
for the spectral shapes during outburst, see e.g. Chen et al 1995; 
Chakrabarti \& Titarchuk 1995; Esin et al 1997 and Chakrabarti 1998.)
If it is a black hole, then that energy is lost!  But if it is a neutron 
star then the energy will be radiated from the neutron star's surface.  
The model therefore predicts that black-hole SXTs will be X-ray 
fainter in quiescence (relative to outburst) than neutron-star systems, 
and there is some evidence for this (see discussion in McClintock 1998).

\subsection {X-ray Spectroscopy of BHXRBs}

With Cyg X-1 as the first BH candidate, its X-ray properties were not
surprisingly proposed as key indicators to help search for similar systems.
Cyg X-1 exhibits 2 X-ray ``states'', a low, hard state with a power law
spectrum extending to very high energies, and a high, soft state where the
low energy data are well represented by a (multi-colour or disk) black-body
spectrum at a temperature of $kT\sim$1~keV.  The power law can extend to
hundreds of keV, sometimes to MeV.  This is usually attributed to
Comptonisation, in which case the highest energy photons require more
scatterings, and hence an energy-dependent time delay would be expected.
Such an effect is seen in Cyg X-1 (see van Paradijs 1998 and references
therein) but it is complex.

The black-body component is explained as arising from the inner accretion
disc around the compact object.  Assuming that this could be approximated by
a ``multi-colour disc'' (MCD) model which incorporates the temperature
variation with disc radius, Mitsuda et al (1984) fitted the Ginga X-ray
spectra to show that all those suspected (on other grounds) of being BHC had
inner disc radii of $\sim$3$r_S$, appropriate for stellar mass black hole
candidates.  $r_S$ is the Schwarzschild radius, and $\sim$3 times this value
is the last stable orbit for matter around such an object. Interestingly,
correcting the MCD model to allow for the effects of GR made no difference
to the fitting of the continuum spectrum.  However, it does affect the
profiles of the spectral lines, and this is believed to have been seen in
AGN (Tanaka et al, 1995; see accompanying article by Madejski), but not yet
in XRBs (see Ebisawa 1997, although
\.{Z}ycki et al (1997) have shown that such profiles are consistent with the
X-ray outburst Fe line spectra obtained by Ginga). It has also recently been
pointed out (Zhang et al 1997) that the BH spin can affect the size of
$r_S$, from $\sim$1--9 times the radius of the event horizon, and this will
be discussed further below.

\subsection{X-ray Spectral/Variability States}

The presence of type-I X-ray bursts or pulsations immediately identifies 
the compact object as a neutron star, and detailed modelling of their 
light curves leads to constraints on the physical properties of the 
compact object.  However, the discovery of rapid but non-periodic 
variability ({\it quasi-periodic oscillations} or QPOs) by EXOSAT in 
GX5-1 and other LMXBs (see Lewin, van Paradijs \& van der Klis 1988) 
opened up new avenues of investigation into the physical processes 
occurring close to the interface between the accretion disc and compact 
object.  Simultaneous analysis of the X-ray spectral and temporal 
variability led to the description of the behaviour of the source in 
terms of source ``states'' which could be understood as functions of the 
mass accretion rate (see review by van Paradijs 1998).

The 2-state behaviour of Cyg X-1 (with its low state, hard X-ray spectrum 
contrasting with the soft, high state) has been well-studied since the 
1970s and is even considered as a possible black-hole diagnostic.
Similar studies of the BHXRBs have extended these concepts and 
shown interesting correlations between 
their X-ray spectra and temporal variability.  The presence of QPOs 
also reinforces their interpretation as a property of the inner accretion disc 
rather than the compact object.  The SXT outbursts, when studied from 
peak to quiescence, cover a very 
large range in accretion rate onto the compact object, and detailed X-ray 
observations of N Mus 1991 by Ginga (van der Klis 1994, 1995; see figure 2) 
suggested that the 
change in its X-ray spectral and temporal behaviour was indeed a simple 
function of mass accretion rate.  Table 2 summarises the key X-ray 
states observed so far in SXTs as defined by the temperature of the X-ray 
spectrum and variability characteristics displayed by the power density 
spectrum (PDS).

\begin{figure}[t]
\begin{center}
\begin{picture}(100,250)(50,30)
\put(0,0){\includegraphics{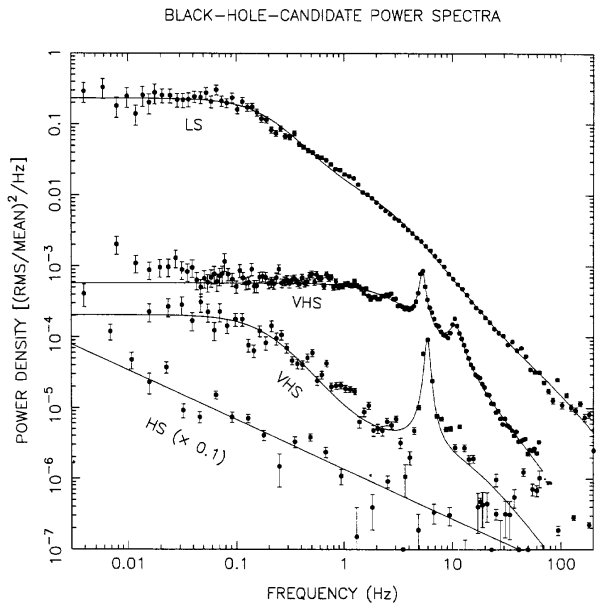}}
\noindent
\end{picture}
\caption[{\bf Figure 2}] {Power spectra of BHXRBs in various spectral states
as observed by {\it Ginga}.  The low state is of Cyg X-1, whereas the others
are of Nova Mus 1991.  From van der Klis (1995).}

\label{fig2}
\end{center}
\end{figure}

\begin{table}
\begin{center}
\caption{X-ray Spectral and Temporal Properties of SXTs}

\begin{tabular}{cll} 
\hline
{\em Source State} & {\em X-ray Spectrum} & {\em Temporal 
Characteristics}  \\
\hline

Low (LS) &  no ultra-soft (US) component & power law PDS, \\
 & & substantial variability  \\
Intermediate (IS) & US + steeper power law at high E & Lorentzian 
noise  \\
High (HS) &  US dominates (MCD model), &  very little variability  \\
 & very weak power law component &  \\
Very high (VHS) & strong US + PL component & strong QPOs at $\sim$
10Hz,  \\
& & Lorentzian noise  \\
\hline

\end{tabular}
\end{center}
\end{table}

However, the fact that the IS has properties similar to the VHS indicate 
that the problem must be more complex than just a function of mass 
transfer rate.  Care must also be taken in drawing comparisons between 
observations made in different energy bands, as can be seen by examining 
the decay light curves of Nova Mus 1991 by BATSE (see Ebisawa et al 1994).

It should also be noted that:

\begin{itemize}

\item the hard power law component in Cyg X-1 is very stable, it is 
the US component 
that can change on timescales of $\sim$ day(s);

\item the US component is {\it anti-}correlated with the radio emission;

\item Barret et al (1996) proposed that observations are consistent with 
only the BHXRBs exhibiting both a hard power law component and a high $L_X$  
(but note the energy range concern);

\item LS BHXRBs and neutron star systems in the atoll state are very 
similar.  Hence 
the presence of a solid surface and strong magnetic field makes little 
difference in this situation! 

\end{itemize}

\subsection{Light Curve Shapes}

The X-ray light curves are reviewed by Tanaka \& Shibazaki (1996), but
are most comprehensively presented by Chen et al (1997).  In spite of the
similarity of GS2023+338, GS2000+25 and N Mus 1991, they do in fact
exhibit a wide variety of outburst behaviour.  Nevertheless the basic
shape is that of a fast rise followed by an exponential decay, the latter
being interpreted as a disc instability (see Cannizzo 1998, King \&
Ritter 1997).  However, more recently 
Shahbaz et al (1998a) have shown that SXTs can exhibit both exponential and
linear decays.  As pointed out by King \& Ritter (1998), the exponential
decay is a natural consequence of the X-ray irradiation maintaining the disc
in a hot (viscous) state, thereby producing outbursts that last much longer
than in their dwarf nova analogues.  If the SXT outburst is not bright
enough to ionise the outer edge of the disc, then a linear decay will
result.  This is shown in figure 3 for those systems with known orbital
period and whose decay light curve shape can be determined.
Furthermore, Shahbaz et al show that the radius of the hot disc at peak of
outburst is related to the time at which the secondary maximum is seen in
the light curve, opening up the possibility of using the light curve as form
of standard candle.

\begin{figure}[t]
\begin{center}
\begin{picture}(100,250)(50,30)
\put(0,0){\includegraphics{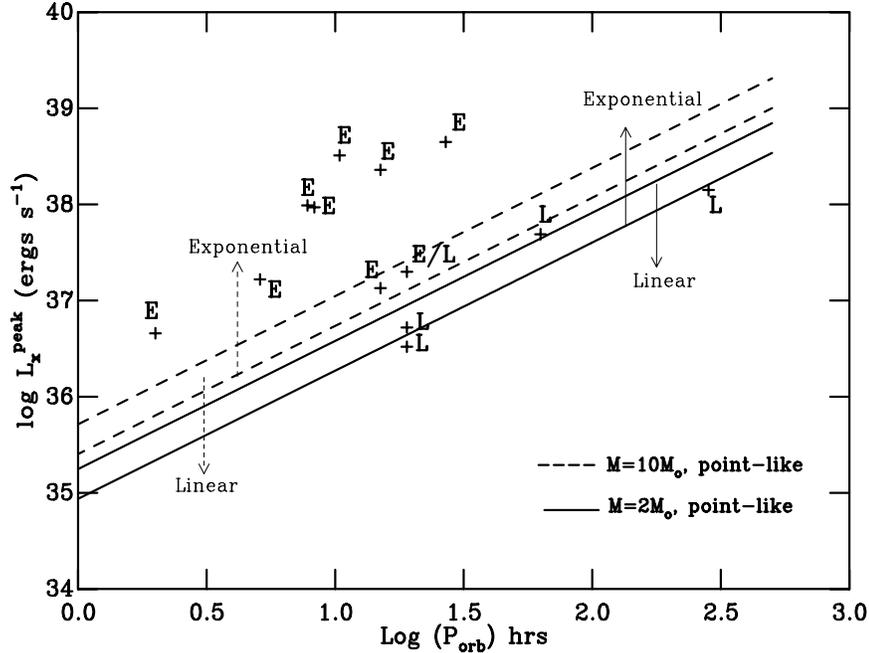}}
\noindent
\end{picture}
\caption[{\bf Figure 3}] {The critical luminosity needed to ionise the entire
accretion disc, according to the calculations of Shahbaz et al 1998a.
Results are shown for total masses of 2 and 10$M_\odot$ corresponding to
neutron star and black hole SXTs respectively.  These luminosities are a
factor 2 smaller for exponential decays compared to linear ones which is due
to the difference in the circularisation and tidal radii of the disc.  The
SXTs shown are SAX J1808.4-3658, GRO J0422+32, A0620-00, GS2000+25,
GS1124-68, Cen X-4, Aql X-1, 4U1543-47, GRO J1655-40 and GRO J1744-28.
Observed decay shapes are designated E (exponential) and L (linear).}

\label{fig3}
\end{center}
\end{figure}

As mentioned earlier Shahbaz \& Kuulkers
(1998) have shown that the optical outburst amplitude ${\Delta}V$ is related
to the system's orbital period (for periods $<$1 day) according to:

\begin{equation}
{\Delta}V = 14.36 - 7.63 \log{P}
\end{equation}

where $P$ is given in hours.  This relation is shown along with the observed
points in figure 4, and is essentially due to the brighter secondaries
(which must be Roche-lobe filling) in longer period systems.  This now
provides the valuable function of being able to predict the orbital period
providing the peak and quiescent magnitudes are known.

\begin{figure}[t]
\begin{center}
\begin{picture}(100,250)(50,30)
\put(0,0){\includegraphics{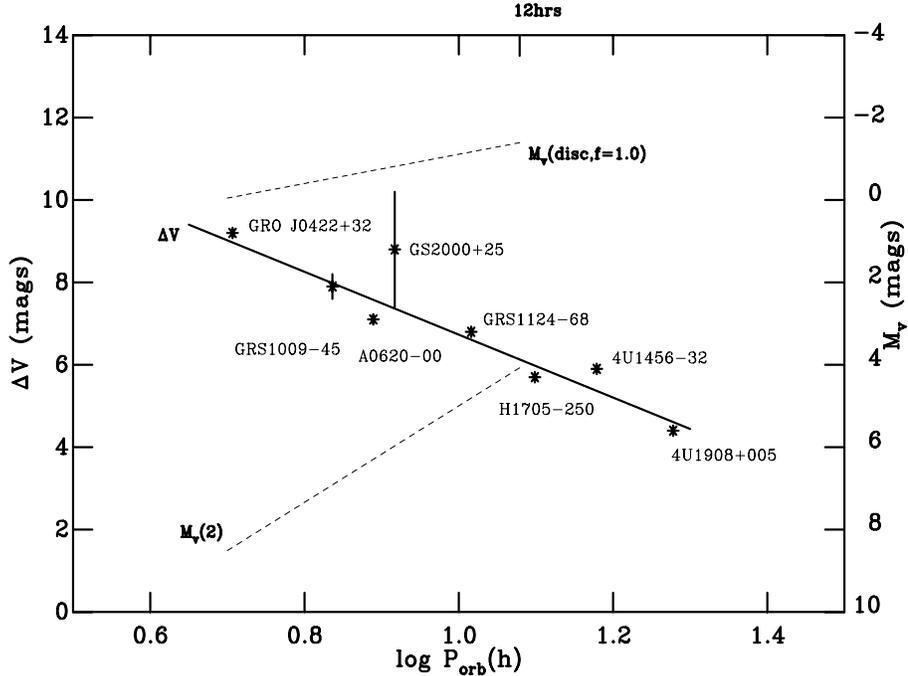}}
\noindent
\end{picture}
\caption[{\bf Figure 4}] {Visual outburst amplitude of SXTs as a function of
orbital period, together with the least-squares fit (taken from Shahbaz \&
Kuulkers 1998).  The dashed lines show
the calculated $M_V$ of the disk in an outbursting SXT and the $M_V$ of the
secondary star according to Warner (1995).}

\label{fig4}
\end{center}
\end{figure}

However, a key observational problem concerns SXTs' quiescent 
properties.  Optical spectroscopy shows, in all cases strong H$\alpha$ 
emission, doppler tomography of which indicates continuing mass transfer 
into the accretion disc.  However, the implied rate is not consistent 
with the very low observed quiescent X-ray fluxes.  This led Narayan et 
al (1997b) to propose ADAFs as a means of accounting for this difference 
between the BH and NS systems.  At low mass transfer rates, the 
temperature and density of gas in the inner disc region would be such as 
to produce a cooling time for the gas that exceeded the radial infall 
time.  Hence the thermal energy of the gas would be advected into the 
black hole, and thereby lost to the observer.  And while similar flows 
would occur in NS systems, the thermal energy would eventually be 
radiated from the NS surface, hence producing an apparently much higher 
$L_X$ from the same rate of accreting matter.

\subsection{Effect of Black Hole Spin}

In spite of their observational title of SXT, there are two transients 
(GS2023+338 and GRO~J0422+32) that do not show US components in their 
bright state, nor do Cyg X-1 and GX339-4 when in their hard (low) state.  
Both optical and X-ray studies show that this cannot be due to a high 
inclination in any of these.  Zhang et al (1997) have calculated the disc 
emission from both Schwarzschild and Kerr BHs, allowing the specific 
angular momentum ($a_* = a/r_g$, where $r_g = GM/c^2$) to take values of 0 (non-spinning hole) 
and $\pm$1 (for maximal spin) where a negative value implies a retrograde 
spin.  From this they inferred that we would observe a black-body (US) 
component whose colour temperature was a function of $a_*$.

Given the great distances of the SXTs (they are rare and spread 
throughout the galaxy), a key requirement of the US component is that it 
has $kT_{col}\geq$0.5--1 keV, in order for the emission not to be 
obscured by interstellar absorption.  The calculations show that the 
highest $kT_{col}$ (and most luminous US component) occur for $a_*=+1$, 
and hence that GRO J1655-40 and GRS1915+10 must both have this value.  And they 
are both jet sources!

Interestingly it can already be asserted that GRO~J1655-40 {\it cannot} be a 
Schwarzschild BH as the dynamically determined $M_X=7M_{\odot}$ requires 
the radius of the last stable orbit to be 
$2.3r_g$ and yet theoretically the minimum value required is $6r_g$.

\section{Mass Measurements}

\subsection{Radial Velocity Curves}

It is when they reach quiescence that the SXTs become such valuable 
resources for research into the nature of LMXBs.  Their optical 
brightness has typically declined by a factor of 100 or more, with all 
the known SXTs having quiescent magnitudes in the range 17--23.  The 
quiescent light is now dominated by the companion star and, while 
technically challenging, presents us with the opportunity to determine 
its spectral type, period and radial velocity curve (whose amplitude is 
the $K$-velocity).    From the latter two 
we can calculate the mass function (equation 1.3)
and the results for the same 9 systems (this time listed in order of 
their mass functions) are summarised in table 3, again separating out the 
two early spectral type systems as well as the single neutron star SXT, 
Cen X-4.  Hence the enormous importance 
of the SXTs, since all are LMXBs which have $M_X>M_2$.  The mass 
functions in table 3 represent the {\it absolute minimum} values for 
$M_X$ since (for all of them) $i<90^o$ and $M_2>0$, both of which 
serve only to {\it increase} the implied value of $M_X$.  That is why the 
work of McClintock \& Remillard (1986) on A0620-00 and Casares et al 
(1992) on V404 Cyg has generated so much interest.

It should also be noted that it can be possible to derive some dynamical 
information about the system even during outburst, providing 
spectroscopic data of sufficient resolution is obtained.  Casares et al 
(1995) observed GRO~J0422+32 during one of its subsequent 
``mini-outbursts'' and found intense Balmer and HeII$\lambda$4686 
emission that was modulated on what was subsequently shown to be the 
orbital period.  Furthermore, a sharp component of HeII displayed an 
S-wave that was likely associated with the hotspot.  

However, to determine 
the actual value of $M_X$ we need additional constraints that will allow 
us to infer values for $M_2$ and $i$. 

\begin{table}
\begin{center}
\caption{Derived Parameters and Dynamical Mass Measurements of SXTs}

\begin{tabular}{lccccccl} 
\hline
{\em Source} &  {\em f(M)} & $\rho$ & 
{\em q} & {\em i} & {\em M$_X$} & {\em M$_2$} & \\
 &  ($M_\odot$) & {\em (g~cm$^{-3}$)} & (=M$_X$/M$_2$) &  & ($M_\odot$) & ($M_\odot$) & Ref. \\  
\hline

V404 Cyg  &  6.08$\pm$0.06 & 0.005 & 17$\pm$1 & 55$\pm$4 & 12$\pm$2 &
0.6 & [1--2] \\
G2000+25  & 5.01$\pm$0.12 & 1.6 & 24$\pm$10 & 56$\pm$15  &
10$\pm$4 & 0.5 & [3--5] \\
N Oph 77  &  4.86$\pm$0.13 & 0.7 & $>$19 & 60$\pm$10 & 6$\pm$2
& 0.3 & [6--9] \\
N Mus 91  &  3.01$\pm$0.15 & 1.0 &  8$\pm$2 &
54$^{+20}_{-15}$ & 6$^{+5}_{-2}$ & 0.8 & [13--15] \\
A0620-00  &  2.91$\pm$0.08 & 1.8 & 15$\pm$1 & 37$\pm$5 & 10$\pm$5 & 0.6 &
[16--18] \\
J0422+32   & 1.21$\pm$0.06 & 4.2 & $>$12 & 20--40 &
10$\pm$5 & 0.3 & [19--20]\\
 & & & & & & & \\
J1655-40  &  3.24$\pm$0.14 & 0.03 & 3.6$\pm$0.9 & 67$\pm$3 & 6.9$\pm$1 & 2.1 &
[10--12] \\
4U1543-47 & 0.22$\pm$0.02 & 0.2 & - & 20--40 & 5.0$\pm$2.5 & 2.5 & [21] \\
 & & & & & & & \\
Cen X-4  &  0.21$\pm$0.08 & 0.5 & 5$\pm$1 & 43$\pm$11 &  1.3$\pm$0.6 & 0.4 &
[22--23] \\
\hline
\end{tabular}
\footnotesize
[1] Casares \& Charles 1994; [2] Shahbaz et al 1994b; [3] Filippenko et 
al 1995a; [4] Beekman et al 1996; [5] Harlaftis et al 1996; 
[6] Filippenko et al 1997; [7] Remillard 
et al 1996; [8] Martin et al 1995; [9] Harlaftis et al 1997; 
[10] Orosz \& Bailyn 1997; [11, 12] van 
der Hooft 1997, 1998; [13] Orosz et al 1996; [14] Casares et al 1997;
[15] Shahbaz et al 1997;
[16] Orosz et al 1994; [17] Marsh et al 1994; [18] Shahbaz et al 1994a; 
[19] Filippenko et al 1995b; [20] Beekman et al 1997; [21] Orosz et al 
1998; [22] McClintock \& Remillard, 1990; [23] Shahbaz et al 1993. 
\normalsize
\end{center}
\end{table}

\subsection{Rotational Broadening}

Since the secondary is constrained to corotate with the primary in short 
period interacting binaries, we can exploit our knowledge of its size by 
making {\it assumption 1} that $R_2$ is given by equation (1).
Hence the result (Wade and Horne, 1988)

\begin{equation}
v_{rot}\sin i = {{2{\pi}R_2}\over{P}}{\sin}i = K_2\times
0.46{(1+q)^{2/3}\over{q}} 
\end{equation}

from which $q$ can be derived if $v_{rot}$ is measurable.  Typical values 
are in the range 40--100 km~s$^{-1}$ and clearly require high resolution 
and high signal-to-noise spectra of the secondary.  

\begin{figure}[t]
\begin{center}
\begin{picture}(100,250)(50,30)
\put(0,0){\includegraphics{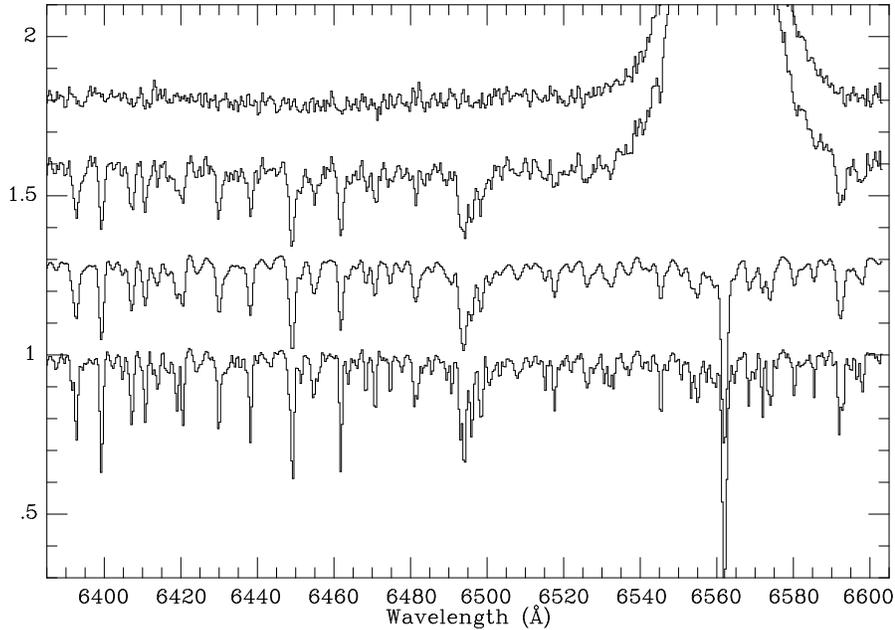}}
\noindent
\end{picture}
\caption[{\bf Figure 5}] {\it Determining the rotational broadening in
V404 Cyg. From bottom to 
top: the K0IV template star (HR8857); the same spectrum broadened by 39 
km s$^{-1}$; doppler corrected sum of V404 Cyg (dominated by intense 
H$\alpha$ emission from the disc); residual spectrum after subtraction of 
the broadened template (from Casares and Charles, 1994).
\label{fig5}}
\end{center}
\end{figure}

Figure 5 (second from 
top) shows the 
Casares \& Charles (1994) WHT summed spectrum of V404 Cyg after doppler 
correcting all individual spectra into the rest-frame of the secondary.  
The bottom spectrum is a very high S/N spectrum of a K0IV star which was 
used as a template, and which clearly has much narrower (actually they 
are unresolved) absorption lines.  The template is broadened by different 
velocities (together with the effects of limb darkening), subtracted from 
that of V404 Cyg and the residuals $\chi^2$ tested.  This gave 
$v_{rot}\sin i = 39\pm1$ km s$^{-1}$
and hence $q$ = 16.7$\pm$1.4.  
The full details can 
be found in Casares \& Charles, and Marsh et al (1994).   It should also 
be noted that while the accretion disc around the compact object might be 
expected to provide some velocity information, there are serious 
difficulties with this.  The H$\alpha$ line in figure 5 is extremely 
broad ($\geq$1000 km~s$^{-1}$) and yet the compact object's motion in 
such high $q$ systems will be very small (typically 
$\leq$30 km~s$^{-1}$).  Nevertheless such motions have been seen (e.g. 
Orosz et al 1994), but their interpretation is not straightforward as 
there is a small phase offset relative to the motion of the companion 
star, and so they cannot be used as part of the dynamical study.

Having 
determined $q$, the range of masses consistent with the observed $f(M)$ 
is plotted in figure 6, where the only remaining unknown is the orbital 
inclination $i$.  To date, none of the SXTs is eclipsing (although 
GRO~J1655-40 shows evidence for a grazing eclipse), and so it is the 
determination of $i$ that leads to the greatest uncertainty in the final 
mass measurement.  Nevertheless there are methods by which $i$ can be 
estimated.

Note also, that high mass ratios for these systems is also implied by the
work of O'Donoghue \& Charles (1996) which demonstrates the {\it superhumps}
have been seen in SXT optical light curves during outburst decay.  Such
features had been seen before during the {\it superoutbursts} of the SU UMa
subclass of dwarf novae and are attributed as arising from tidal stressing
of the accretion discs in high mass ratio interacting binaries.  Their
calculation of system parameters based on this model provides satisfactory
confirmation of these values.

\begin{figure}[bt]
\begin{center}
\begin{picture}(100,250)(50,30)
\put(0,0){\includegraphics{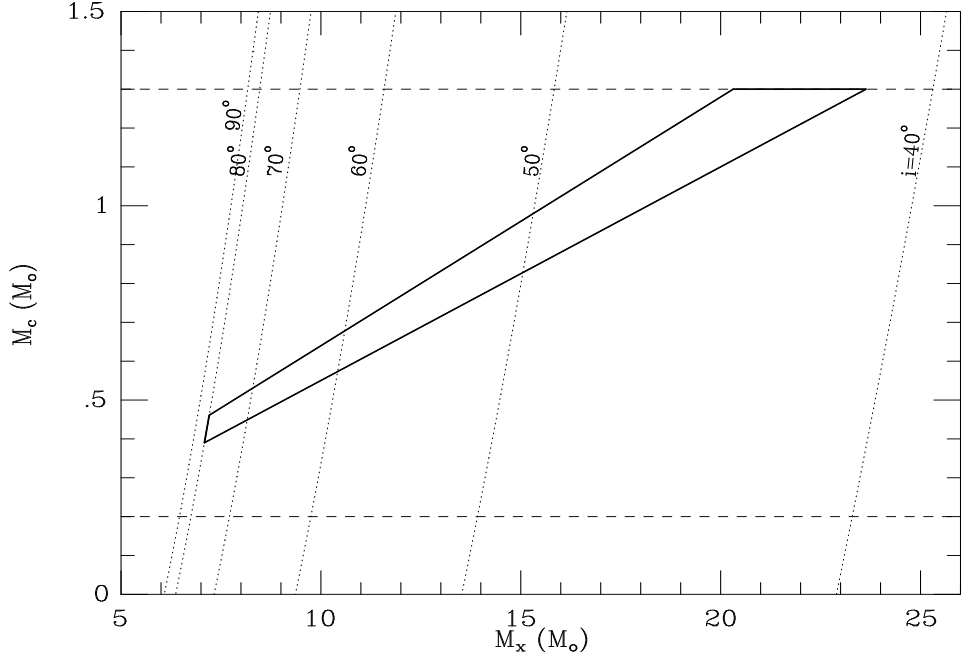}}
\noindent
\end{picture}
\caption[{\bf Figure 6.}] {\it Constraints on $M_X$ and $M_2$ for a range
of values of $i$ in V404 Cyg based on the radial 
velocity curve ($f(M)$) and determination of $q$ (from the rotational 
broadening).  It is the limited constraint on $i$ (absence of eclipses) 
that leads to a wide range of $M_X$ (from Casares and Charles, 1994).
\label{fig6}}
\end{center}
\end{figure}

\subsection{Ellipsoidal Modulation}

We exploit one more property of the secondary, it's peculiarly distorted
shape responsible for the so-called {\it ellipsoidal modulation} as we view
the varying projected area of the secondary around the orbit.  This leads to
the classical double-humped light curve, as shown for A0620-00 in figure 7.
If the secondary's shape is sufficiently well-determined by theory (i.e. the
form of the Roche lobe) then the observed light curve depends on only 2
parameters, $q$ and $i$.  In several cases (as described in the previous
section) $q$ is already determined, but in practice the ellipsoidal
modulation is largely insensitive to $q$ for values $q>$5.  Details of the
light curve modelling can be found in Shahbaz et al (1993), and the
collected results are in table 3.

\begin{figure}[t]
\begin{center}
\begin{picture}(100,250)(50,30)
\put(0,0){\includegraphics{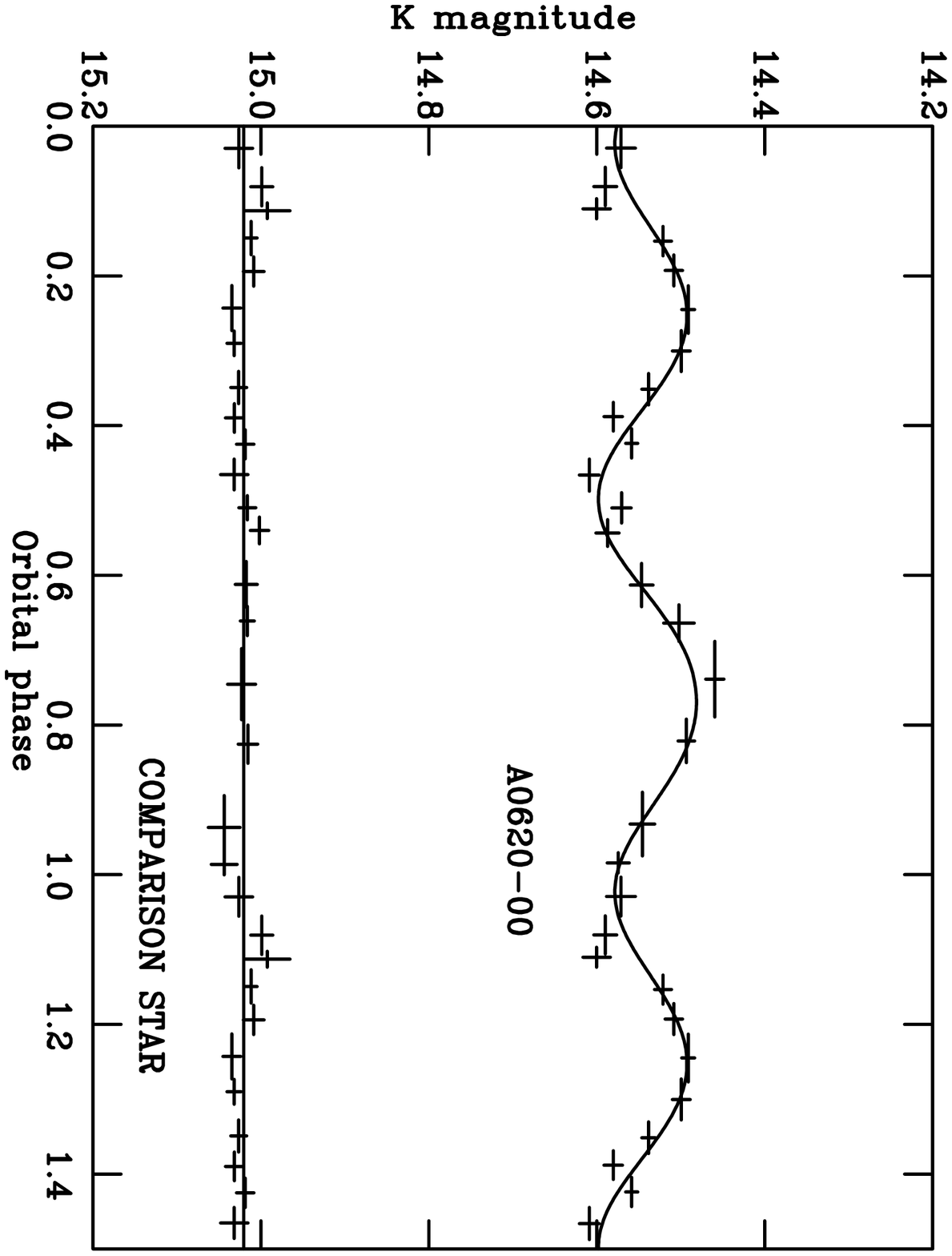}}
\noindent
\end{picture}
\caption[{\bf Figure 7}] {\it IR light curve of A0620-00 showing the
classical double-humped modulation (from Shahbaz et 
al, 1994a).
\label{fig7}}
\end{center}
\end{figure}

This final stage in the SXT orbital solutions has made 2 key assumptions:
{\it assumption 2}, that the secondary in quiescence fills its Roche lobe;
{\it assumption 3}, that the light curve is not contaminated by any other
light sources.  It is felt that the former is reasonable since there is
strong evidence through doppler tomography (e.g. Marsh et al 1994) for
continued mass transfer in quiescence from the secondary. However, the
principal (and potentially significant) uncertainty is the problem of any
other contaminating light sources.  This would mainly be the accretion disk,
but residual X-ray heating and starspots on the surface of the secondary
might also be present.  It is for this reason that this work has been
performed in the K band whenever possible.  The disc contamination has been
measured in the optical around H$\alpha$ (as a by-product of the spectral
type determination by searching for excess continuum light) and is typically
$\leq$10\%.  It should therefore be even less in the IR given the blue
colour of the disc. However, the outer disc edge has been found to be an IR
emitter in CVs (Berriman et al 1985) and the light curves might be
contaminated, as was suggested in the case of V404 Cyg by Sanwal et al
(1996).  This is potentially an important effect, since a contaminating (and
presumably steady) contribution will reduce the amplitude of the ellipsoidal
modulation, which will lead to a lower value of $i$ being inferred, and
hence a higher mass for the compact object.

For this reason Shahbaz et al (1996, 1998b) undertook IR K-band 
spectroscopy of the two brightest and best studied SXTs, V404 Cyg and 
A0620-00.  Only upper limits were derived in both cases, showing that any 
contamination must be small and hence the masses derived can (at most) be 
reduced by only small amounts.  It is also interesting to note that, in 
their study of the non-orbital optical variability in V404 Cyg, Pavlenko 
et al (1996) found that (as first noted by Wagner et al 1992) the 
ellipsoidal modulation could be discerned underlying the substantial 
flickering in the light curve.  Interpreting the 
flickering as a completely independent component, Pavlenko et al showed 
that the {\it lower envelope} of this light curve (rather than the mean) 
produced an ellipsoidal 
light curve which, when fitted as described above, gave essentially 
identical results to those obtained from the K-band analysis, thereby 
providing further weight to the significance of the mass determinations. 

The results of these analyses are collected together in figure 8, which
contains all neutron star and black hole mass measurements.  It should also
be noted that the value for Cen X-4 (one of only two neutron star SXTs,
identified on the basis of its type I X-ray bursts) has been derived exactly
by the method outlined here (Shahbaz et al 1993) and yields a value of
1.3M$_\odot$, in excellent accord with that expected for a neutron star.

\begin{figure}[t]
\begin{center}
\begin{picture}(100,250)(50,30)
\put(0,0){\includegraphics{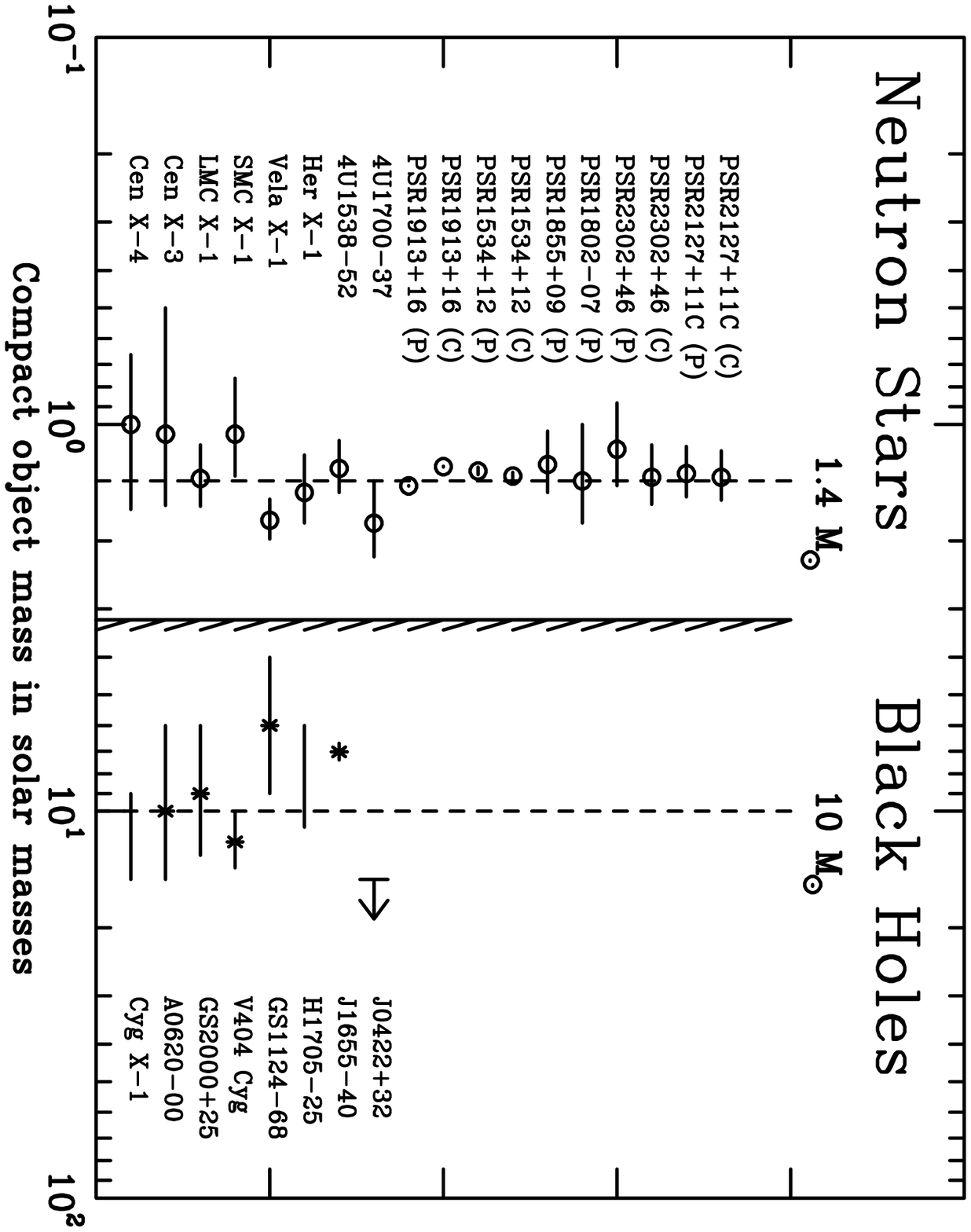}}
\noindent
\end{picture}
\caption[{\bf Figure 8}] {The mass distribution of neutron stars and black 
holes.  Note the
remarkably narrow spread of neutron star masses, and the large factor by
which the BHXRB masses exceed the (canonical) maximum mass of 3.2$M_\odot$.}
\label{fig8}
\end{center}
\end{figure}

\section{Lithium in the Companion Stars}

One of the remarkable by-products of our high resolution radial velocity
study of V404 Cyg was the discovery of strong LiI $\lambda$6707 absorption
in the secondary star (Mart\'\i{}n et al 1992).  This was, of course, not
present in any of the template stars which we were using for the spectral
fitting, since Li is a characteristic feature of young, pre-main sequence
and T Tau objects.  Subsequent convection in late type stars leads to the
destruction of lithium, and observed abundances in normal stars (like the
Sun) are a thousand times lower.  Similar Li enhancements have since been
identified in A0620-00, J0422+32, GS2000+25, N Mus 1991 and Cen X-4 (see 
Mart\'\i{}n et al 1996), the latter indicating that this phenomenon
is not confined solely to the BHC.

Li is an important element in galactic chemical abundances because galactic
material is found to be enriched in Li relative to the halo, and the source
of this enrichment is a subject of current research.  Since the SXTs are
clearly highly evolved objects which are extremely unlikely to have retained
such high Li abundances, then we must find mechanisms create Li within the
structure of an SXT.  Those systems in which Li has been detected cover a
wide range of period and secondary size, but they all display high
luminosity, recurrent X-ray outbursts.  Mart\'\i{}n et al (1994) therefore
suggested that spallation processes during these outbursts produce Li in
large quantities close to the compact object.  Subsequent large mass
outflows result in the transfer of some of this Li to the secondary, the
energetics of which are considered by Mart\'\i{}n et al (1994).  Support for
this suggestion is cited as the interpretation of the 476keV $\gamma$-ray
line that was observed during the N Mus 1991 outburst (Sunyaev et al 1992;
Chen et al 1993).  Originally interpreted as the gravitationally redshifted
$e^-$-$e^+$ 511keV line, it might instead be associated with the 478keV line
of $^7$Li.  The temporal behaviour of this line (it only lasted for about 12
hours) could give insight into the spallation mechanism.  That high
luminosities are needed, whatever the mechanism, is indicated by the absence
of Li in CVs of comparable spectral type to the SXTs, nor has Li been
detected in the nova GK Per (Mart\'\i{}n et al 1995).

\section{The Superluminal Transients}

In 1994 there were two new X-ray transients discovered, GRS1915+105 and 
GRO~J1655-40, that brought an entirely new type of behaviour to this field.  As
with many of the transient outbursts, they also emitted strongly in the
radio, but VLA and VLBI observations showed that these objects also
exhibited ejection events that were ``superluminal'' 
(Mirabel \& Rodr\'\i{}guez
1994; Hjellming \& Rupen 1995), the first time that such phenomena had been
observed within the Galaxy.  Further dynamical studies of GRS1915+105 are 
severely hampered by (a) its extremely high interstellar extinction 
($A_V{\sim}$26), leaving only a variable, K$\sim$14 IR counterpart,
and (b) its continuing and extremely variable X-ray 
activity that is totally unlike any of the ``classical'' SXTs.  It is not 
even clear that GRS1915+105 is
an LMXB (see Mirabel et al 1997), and it demonstrates an
extraordinarily rich variety of X-ray variability (e.g. Morgan \& Remillard 1996,
Belloni et al 1997).  

GRO~J1655-40 (N Sco 1994), on the other hand, is optically the brightest 
in quiescence of all the SXTs, and so has extremely well-determined 
photometric light-curves, and is an excellent candidate for a dynamical 
study.  The companion also has one of 
the earliest (confirmed) spectral types (mid-F) of the SXTs which means 
that, in quiescence, the effects
of the accretion disk are very small, almost negligible.  And the high
$\gamma$-velocity led to the suggestion (Brandt et al 1995) that J1655-40
could be an example of a NS system that had suffered accretion-induced
collapse.  However, J1655-40's behaviour is not at all typical of other 
SXTs, with quiescent studies severely hampered by its return to activity 
in 1996.  This return was fortuitously observed by Orosz et al (1997) who 
found that the optical brightening began $\sim$6 days before the X-ray 
activity began.  They (and Hameury et al 1997) interpreted this as an 
``outside-in'' outburst of the accretion disc, with the substantial delay 
arising due to the ADAF flow (in quiescence) having evaporated the inner 
disc, and which needed to be re-filled before accretion onto the compact 
object could take place.  Subsequent multi-wavelength (UV/optical/X-ray 
with HST and RXTE) observations of 
this period of activity (Hynes et al 1998) demonstrated two interesting 
properties.  They found that the X-ray and optical variations were, at 
times, correlated, but with the optical variations lagging the X-ray by 
$\sim$19 secs.  With the known size of the binary from its orbital 
period, this lag is too short to be due to irradiation of the secondary, 
and hence must be associated with the accretion disc.  Furthermore, Hynes 
et al found that as the outburst progressed, the optical/UV emission 
declined as the X-rays increased!  They suggested that this might arise 
through the driving of a large corona early in the outburst which can 
then allow subsequent up-scattering of hard X-rays later on.   However, 
this requires much more extensive and detailed multi-wavelength studies 
to be undertaken throughout an outburst in order for it to be fully 
tested.

The orbital system parameters have been derived from several photometric
studies of J1655-40 by van der Hooft et al (1997), Orosz \& Bailyn (1997)
and van der Hooft et al (1998).  Figure 9 shows the van der Hooft et al
light curve together with the system schematic of Orosz \& Bailyn.  
The values recorded in table 3 are those
from the latter paper due to their more conservative error analysis.
J1655-40 is unusual in this class in that it has a 
low mass ratio of $q{\sim}3$ (but this has not
yet been obtained from a rotational broadening study, due to its return to
activity shortly after its initial outburst).  At such a value, the
ellipsoidal modulation is sensitive to both $q$ and $i$ (the latter also
being tightly constrained here as a result of its grazing eclipse).  Hence,
once it becomes possible to spectroscopically determine the rotational
broadening of the F star (when it re-enters an extended period of
quiescence) it will then be possible to perform a check of the entire basis
on which the quiescent SXT light curves have been modelled and used to
determine $i$. It has also been suggested (Kolb et al 1997) that the
secondary star is in a very interesting evolutionary state in which it is
crossing the Hertzsprung gap and about to ascend the giant branch.  This is
what is driving the much higher mass transfer rate than in the other SXTs,
but temporary drops in \.{M} return it to the transient domain.

\begin{figure}[t]
\begin{center}
\begin{picture}(100,250)(50,30)
\put(0,0){\includegraphics{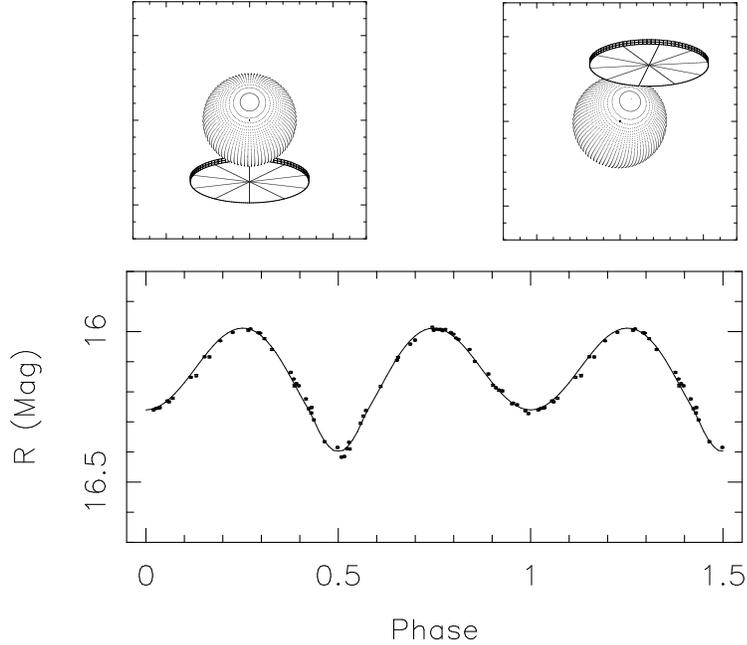}}
\noindent
\end{picture}
\caption[{\bf Figure 9}] {R-band light curve of GRO J1655-40 taken by van
der Hooft et al (1997) when the source was quiescent.  To obtain acceptable
fits (shown) it was necessary to add partial obscuration by a cool accretion
disc at phase 0.5, as demonstrated in the system schematic (above) taken
from Orosz \& Bailyn (1997).}

\label{fig9}
\end{center}
\end{figure}

\section{Outburst Mechanisms}

Over the last 10 years there has been the same debate over the 
mechanism for SXT outbursts as had been taking place over the cause of 
dwarf nova outbursts, with
the same two competing models, namely enhanced mass transfer from the secondary
star (as a result of X-ray heating) and the thermal (viscous) instability in
the accretion disc itself (both are discussed by Lasota 1996).  However, as
a result of much more sensitive quiescent X-ray observations of SXTs by
ROSAT (Verbunt 1996) it is clear that these levels of X-ray emission (as low
as 2.5$\times$10$^{30}$erg s$^{-1}$ for A0620-00) are incapable of heating
the secondary star sufficiently to generate the mass transfer necessary to
account for the observed outbursts.  And whilst there have been models
proposed (e.g. Chen et al 1993; Augusteijn et al 1993) that combine elements of
both the mass transfer and disc instability explanations, these have not yet
been supported by observations.

Strong support for the disc instability model has appeared in papers
by van Paradijs (1996) and King et al (1996).  They point out that, when
calculating whether the disc instability mechanism will occur in an SXT disc
(which requires that the temperature somewhere in the disc be below the
ionisation temperature of hydrogen, 6500K), it is necessary to take into
account the effects of (time-averaged over outbursts) X-ray heating.  In
this way they derive an expression for the X-ray luminosity (as a function
of period) that separates the steady and transient sources:

\begin{equation}
{\log}L_X = 35.8 + 1.07{\log}P
\end{equation}

and find that this is very well supported by the observations.  Indeed, the
continued activity of GRO J1655-40 has brought it very close to this line,
indicating that it is in fact ``almost'' a steady source.  The average SXT
mass transfer rate from the secondary is found to be:

\begin{equation}
<M_2^.> = -4{\times}10^{-10}P_d^{0.93}M_2^{1.47} M_{\odot}y^{-1}
\end{equation}

giving values $\sim$10$^{-10}$M$_\odot$y$^{-1}$ for most SXTs, but about
3$\times$ higher for GRO J1655-40.

\section{Population Size and Distribution}

Important questions for current and future surveys for X-ray transients,
such as those likely to be undertaken with AXAF and XMM in nearby galaxies
(M31 and M33), concerns the number of quiescent systems, the chance of 
their being observed if they do outburst, and their distribution through
the galaxy.  The largest uncertainty in such calculations is the typical
recurrence time of an SXT.  Following Tanaka \& Shibazaki (1996), 
the current sample's outburst properties 
(table 1) show that this can range from less than a year to $\sim$50 years.   
However, the average is likely to be around 10--20 years, with 
typically 2 SXTs being observed per year (with regular all-sky coverage 
by CGRO and RXTE).  With $\sim$90\% of SXTs located within a galactic 
longitude range of $\pm$80$^o$, this implies that they lie within 
8kpc of the Galactic Centre, but distance estimates for the current 
sample suggest they are all $<$5kpc from us.  Hence, we are only 
detecting $\sim$10\% of the transients that occur within our galaxy (due 
to a combination of interstellar absorption and sensitivity), and 
the total number of SXTs is $\sim$200--1000 for assumed recurrence times 
of 10 and 50 years respectively.  Long-term surveys of nearby galaxies 
will allow these estimates to be more rigorously examined, as well as 
deriving more accurate luminosities and galactic distribution 
information. 

White \& van Paradijs (1996) have examined and compared the distribution in
galactic latitude and longitude of the BHC LMXBs and the NS systems.  They
find that the BHC have a dispersion in $z$ of $\sim$400pc, whereas the NS
are $\sim$1kpc, interpreting the difference as an indication that the kick
velocities received in BH formation are less than in NS.  The corrolary of
this is that the BHC LMXBs are not formed by the accretion-induced collapse
of a NS (except possibly GRO J1655-40) where the higher NS kick would be
inherited by the resultant BH.

\section{Nature of the Compact Object}

Since the existence of compact objects with masses $\geq$10$M_\odot$ can now
be taken as secure, the question arises as to what are they.  Or, more
importantly, what is the maximum mass of a neutron star?  This is usually
quoted as 3.2$M_\odot$ on the basis of the {\it Rhoades-Ruffini
Theorem} (1974).  But there are a number of assumptions built into this theorem
that, if relaxed, can lead to a very different result (see Miller 1997 and
references therein).

In particular, the assumption of causality (which requires the sound speed
to be less than $c$) is only applicable in a non-dispersive medium.  Also,
the density up to which the equation of state is well-defined may be
optimistically high.  Both these effects, together with significant rotation
of the compact object, can lead to the maximum mass of the neutron star only
being constrained to be $<$14$M_\odot$!  Nevertheless, it should be
recognised that current models of neutron star equations of state are
compatible with the original 3.2$M_\odot$ limit.

However, an alternative suggestion that such compact objects may be
$Q$-stars has been made by Bahcall et al (1990).  In such objects, the
strong force confines neutrons and protons at densities below nuclear
density, leading to a very different equation of state (in which the $Q$
stands for conserved $Q$uantity, the baryon number).  They can be very
compact and hence consistent with our current understanding of the
properties of neutron stars.  But Miller, Shahbaz \& Nolan (1998) have shown
that if this model is to be applied to V404 Cyg (i.e. that it is a $Q$-star
of 12$M_\odot$), then it requires that the threshold density for this effect
must be $\sim$10$\times$ {\bf below} that of nuclear density, which is
considered to be too implausible given the results of current experiments.
In which case, we conclude that V404 Cyg and related objects must be black
holes.

\section{Conclusions}

The dramatic advances in the field of galactic black-hole studies have come
about during this decade for 2 main reasons: (i) the almost continuous
monitoring of the X-ray sky that is now provided by all-sky monitors such as
those on CGRO and RXTE has provided a steady stream of new X-ray transients
for subsequent ground-based observations once they reach quiescence, and
(ii) the availability of high performance optical spectrographs with good
red sensitivity on 4m and larger telescopes.  With these facilities we have
obtained more detailed information about the nature of both components of
LMXBs than was hitherto possible.  In particular, the discovery of the mass
function of V404 Cyg has revolutionised attitudes concerning the existence
of compact objects that must be heavier than the canonical maximum mass of a
neutron star.  And while there are useful indicators from X-ray observations
as to the possible presence of black holes in X-ray binaries, the ultimate
diagnostic has to be the dynamical study that has been described here.  The
next major advances will come from observing the many quiescent
transients that are too faint for current 4m class telescopes and require
access to the about to be completed VLT and Gemini telescopes in the 
southern hemisphere.

\section{Future Work}

The SXTs are providing a very fertile hunting ground for BH candidates,
and with CGRO and RXTE both having all-sky monitors there should be a 
steady stream of new transients discovered over the next decade.  Key 
questions to be addressed include:

\begin{itemize}

\item are there any {\it low-mass} ($<$5M$_\odot$) BHs?\\
i.e. formed in 2 stages via accretion-induced neutron star collapse (see
Brown et al. 1996).

\item or any very {\it high-mass} ($>$20M$_\odot$) BHs?\\
i.e. from massive stars that succeeded in retaining most of their mass until
the time of collapse.

\item or are they all formed from He stars at the end of the W-R phase?

\item therefore we need at least a dozen {\it accurate, i.e. $<$10\%} mass
determinations, which will require exploiting the new generation of
8--16m class telescopes.

\item can advective flows demonstrate the existence of the Event Horizon
in the BHXRBs?

\item or high resolution X-ray spectroscopy of Fe emission line profiles
reveal the distorted shape expected due to General Relativity?
\end{itemize}

\bigskip

The observations have now securely established compact objects with masses
in the 5--15M$_\odot$ range, and if they are {\it not} BHs then this has
profound implications for particle physics (and general relativity).  Once
thought impossible, SXTs allow perhaps the cleanest study of the BH
environment.

\begin{acknowledgments}
I am particularly grateful to Erik Kuulkers and Tariq Shahbaz for their 
assistance in the preparation of the figures for this manuscript.
\end{acknowledgments}

\end{document}